\documentclass[12pt]{iopart}
\usepackage{iopams}
\usepackage{graphicx} 

\begin{document}

\title{Magnetism and superconductivity in Eu$_{0.2}$Sr$_{0.8}$(Fe$_{0.86}$Co$_{0.14}$)$_{2}$As$_{2}$ probed by $^{75}$As NMR}
\author[]{R Sarkar \footnote{Author to whom all
 correspondence should be made.}, \dag\ R Nath, P
 Khuntia, \S\ H S Jeevan, \S\ P Gegenwart and M
 Baenitz}
\address{Max-Planck Institute for Chemical Physics of Solids,
01187 Dresden, Germany}
\address{\dag {School of Physics, Indian Institute of Science Education and Research, Thiruvananthapuram-695016, India}}
\address{\S {I. Physik. Institut, Georg-August-Universit$\ddot{a}$t G\"{o}ttingen, 37077 G\"{o}ttingen, Germany }}
\ead{rajibsarkarsinp@gmail.com, rajib.sarkar@cpfs.mpg.de}
\date{\today }

\begin{abstract}

We report bulk superconductivity (SC) in
Eu$_{0.2}$Sr$_{0.8}$(Fe$_{0.86}$Co$_{0.14}$)$_{2}$As$_{2}$ single
crystals by means of electrical resistivity, magnetic
susceptibility, and specific heat measurements with
$T$$_{\mathrm{c}}$ $\simeq$ 20 K with an antiferromagnetic (AFM)
ordering of Eu$^{2+}$ moments at $T$$_{\mathrm{N}}$ $\simeq$ 2.0 K
in zero field. $^{75}$As NMR experiments have been performed in
the two external field directions (H$\|ab$) and (H$\|c$).
$^{75}$As-NMR spectra are analyzed in terms of first order
quadrupolar interaction. Spin-lattice relaxation rates (1/$T_{1}$)
follow a $T^{3}$ law in the temperature range 4.2-15 K. There is
no signature of Hebel-Slichter coherence peak just below the SC
transition indicating a non $s$-wave or s$_{\pm}$ type of
superconductivity. The increase of 1/$T_{1}T$ with lowering the
temperature in the range 160-18 K following $\frac{C}{T+\theta}$
law reflecting 2D AFM spin fluctuations.
\end{abstract}


\maketitle

\section{\textbf{Introduction}}
The recent discovery of superconductivity (SC) in Fe based
pnictides has attracted considerable attention in the condensed
matter physics community to understand the microscopic origin of
the SC and its relation to the Fe based magnetism
\cite{{Kamihara2008},{Chen2008},
{Ren2008},{ChenLi2008},{Yang2008},{Bos2008},{Rotter2010}}. In this
diverse branch of pnictides one family of materials
AFe$_{2}$As$_{2}$(A=Ca, Sr, Ba, Eu) (abbreviated as 122 series)
with the crystal structure of ThCr$_{2}$Si$_{2}$ type exhibits the
SC with the transition temperatures $T_{\rm C}$'s  as high as 38
K\cite{Jeevan2008,Sefat2008,Jasper2008}. Interestingly in this 122
family, EuFe$_{2}$As$_{2}$ is the only member which has a rare
earth moment Eu$^{2+}$ (S=7/2) corresponding to a theoretical
effective moment of 7.94 $\mu_B$. Here the antiferromagnetic
(AFM) ordering of Eu$^{2+}$ moments take place at 19 K and Fe
order AFM (SDW type) at 190 K, which is the highest reported SDW
transition temperature among the pnictide family
\cite{Raffius1993,Jeevan2008a,Wu2009}. The SC could be found in
this compound with the suppression of the Fe ordering by the K
substitution (hole doping) at the Eu$^{2+}$ site with $T_{\rm
C}$'s upto 32 K \cite{Jeevan2008}. In addition to that SC has been
observed in chemically pressurized
EuFe$_{2}$(As$_{1-x}$P$_x$)$_{2}$ alloys accompanied by the
Eu$^{2+}$ ordering \cite{Ren2009,Jeevan2010}.

 In contrast to the other 122 systems, in EuFe$_{2}$As$_{2}$, on
suppression of the AF order of iron with pressure or Co doping,
the onset of SC was observed, however seem to be hindered to
reach zero resistivity because of the magnetic ordering of
Eu$^{2+}$ \cite{{Miclea-PRB-79},{Nicklas-JPCM-273},{Zheng2009}}.
Nevertheless by substituting 80 $\%$ Sr at the Eu$^{2+}$ and
optimal Co doping at the Fe site one can suppress the AFM ordering
of Eu$^{2+}$, and Fe SDW ordering, respectively. Eventually, SC
with the $T_{\rm C}$'s 20 K is being observed
\cite{Rongwei-83-2011}. Apart from the SC, this system could be
also interesting to study the interplay between Fe-$3d$ and
Eu-$4f$ magnetism. Especially if there is any coupling present
between the two subsystem (Fe-$3d$ and Eu-$4f$). Here it is worth
to mention the Eu$^{2+}$ ordering at 19 K in EuFe$_{2}$As$_{2}$
perhaps makes this system much more interesting, because this
could additionally opens up the opportunity to clarify the
mechanism of the interplay between the $3d$ and $4f$ magnetism
and also the influence of Eu$^{2+}$ magnetism on the SC. It has
been already suggested from different experiments that there is
coupling between the localized Eu$^{2+}$ moments and the
conduction electrons of the two dimensional Fe$_{2}$As$_{2}$
layers \cite{{Ren-79-2009},{Jiang-11-2009},{Dengler-81-2010}}.

\indent In this paper we present the electrical resistivity
($\rho$), magnetic susceptibility ($\chi$), and heat capacity
($C_{\rm p}$) measurements on superconducting
Eu$_{0.2}$Sr$_{0.8}$(Fe$_{0.86}$Co$_{0.14}$)$_{2}$As$_{2}$
(abbreviated as ESFCA) single crystal. All these experiments are
complemented by $^{75}$As NMR measurements to shed light on the
microscopic properties. Here it is worth emphasizing numerous NMR
investigations have been done to understand the microscopic
mechanism of the 122 (Ca, Sr, Ba) based superconductors, for
instance
[Ref\cite{Ning2009,Ning2010,Laplace-80-2009,Baek-102-2009,Kitagawa-103-2009}].
However, to the best of our knowledge there is only one reported
NMR investigation on this Eu based 122 system, namely
EuFe$_{1.9}$Co$_{0.1}$As$_{2}$. Guguchia et. al. have suggested,
there is strong coupling between the Eu$^{2+}$ moments and the
Fe$_{1.9}$Co$_{0.1}$As$_{2}$ layers \cite{Guguchia-83-2011}.
However, authors have not focused on the superconducting region.
Therefore still there is a scope to investigate the Eu based
pnictides superconductors to simplify the superconducting
mechanism and the possible role of Eu magnetism on the
supeconducting state. We believe our present investigations will
answer this questions adequately.

\section{\textbf{Experimental details}}
Single crystals of Eu$_{0.2}$Sr$_{0.8}$Fe$_{2}$As$_{2}$ and ESFCA
were synthesized using Sn flux method. Nominal stoichiometric
amounts (0.8:0.2:2:2 for Sr$_{0.8}$Eu$_{0.2}$Fe$_{2}$As$_{2}$ and
0.8:0.2:1.4:0.6:2 for ESFCA) of the respective starting elements
were taken in a alumina crucibles, sealed in a quartz tube under
Ar atmosphere. The batches were heated to 1300 $^{\circ}$C at a
rate of 100 $^{\circ}$C/h and stayed for 6h and then cooled to 900
$^{\circ}$C with the rate 3$^{\circ}$C/h. Crystals were extracted
by etching in diluted HCl acid. The stoichiometry of a
representative crystals was confirmed by semiquantitative
energy-dispersive x-ray (EDX) microanalysis. Zero field cooled
(ZFC) and field cooled (FC) magnetic susceptibility ($\chi$) as a
function of temperature was measured using a commercial Quantum
Design SQUID magnetometer with field applied along the $c$
direction ($H\|c$) and in the $ab$ plane ($H\|ab$) of the
crystal. Temperature dependent $\rho$(T) and $C_{\rm p}$(T)
measurements were performed using a Quantum Design Physical
Property Measurement System (PPMS) for $H\|ab$ only. All the
above measurements were carried out down to 1.8~K except that
$C_{\rm p}$(T) was measured down to 0.37~K by using an additional
$^{3}$He cooling system. The NMR measurements were carried out
using the conventional pulsed NMR technique on $^{75}$As (nuclear
spin $I=3/2$ and gyromagnetic ratio $\gamma/2\pi=7.2914$ MHz/T)
nuclei in a temperature range $4$~K $\leq$ $T$ $\leq $ $160$~K.
The measurements were done at a radio frequency of 48~MHz in
single crystals with external magnetic field H applied along the c
direction ($H\|c$) and within the plane ($H\|ab$). For this
purpose we have taken five single crystals and glued them on top
of each other to make a stack of single crystals. The field sweep
NMR spectra were obtained by integrating the spin-echo in the
time domain and plotting the resulting intensity as a function of
the field. The spin lattice relaxation rate ($1/T_{1}$)
measurements were also performed at 48 MHz in both the field
directions ($H\|c$ and $H\|ab$) following the standard saturation
recovery method by exciting the central transition of $^{75}$As
spectra.

\section{\textbf{Results and Discussion}}
\subsection{\label{sec:level2}Electrical Resistivity, Magnetic Susceptibility and Specific Heat}

\begin{figure}
\includegraphics[scale=0.8]{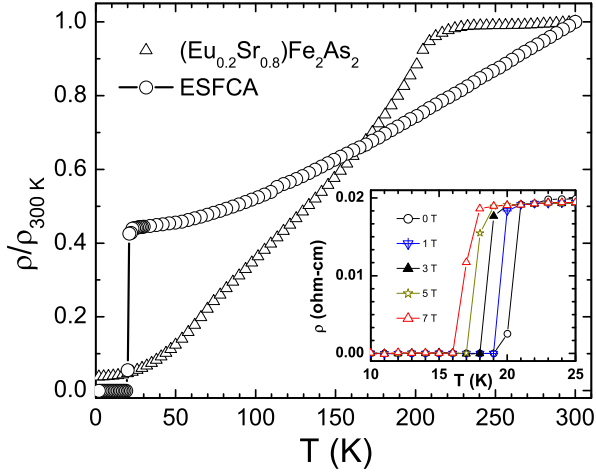}
\caption{\label{resistivity} Temperature dependence of electrical
resistivity $\rho$(T) of Eu$_{0.2}$Sr$_{0.8}$Fe$_{2}$As$_{2}$ and
superconducting ESFCA single crystals. Inset shows the low
temperature $\rho$(T) measured at different applied fields to
highlight the $T_{C}$ of ESFCA. }
\end{figure}

The results of the electrical resistivity normalized by the room
temperature value ($\rho/\rho_{300 K}$(T)) as a function of
temperature is presented in Fig.~\ref{resistivity} for
(Eu$_{0.2}$Sr$_{0.8}$)Fe$_{2}$As$_{2}$ and ESFCA crystals.
$\rho$(T) for (Eu$_{0.2}$Sr$_{0.8}$)Fe$_{2}$As$_{2}$ is weakly
temperature dependent at high temperatures, shows a broad hump at
$T \simeq$ 210~K, and then decreases almost linearly with
decrease in temperature for $T < 210$~K before it attains a
saturation value at $T \leq$~20~K. The broad hump at 210~K is an
indication of the SDW transition which is about 15~K higher than
the transition temperature reported for the parent compound
EuFe$_{2}$As$_{2}$ \cite{Jeevan2008a}. The linear decrease of
$\rho$(T) points towards the metallic nature of the compound. No
trace of superconductivity or Eu$^{2+}$ordering was found down to
2~K suggesting that the Eu$^{2+}$ ordering at 19~K reported for
EuFe$_{2}$As$_{2}$ is suppressed below 2~K after 80\% Sr doping
at the Eu site. The overall behaviour of $\rho$(T) is identical
to the previous report on the parent compound
EuFe$_{2}$As$_{2}$\cite{Jeevan2008a} except that the SDW
transition temperature is enhanced and the Eu$^{2+}$ ordering is
suppressed. At 300~K and 2~K the values of $\rho$ are about
16~m$\Omega$-cm and 0.59~m$\Omega$-cm, respectively yielding a
residual resistivity ratio ($\rho_{300~\rm K}/\rho_{2~\rm K}$) of
27. Such a high value of $\rho_{300~\rm K}/\rho_{2~\rm K}$ is
indicative of a high quality sample with only small amount of
disorder present in the material. As shown in
Fig.~\ref{resistivity} for 14\% Co substitution at the Fe site,
the SDW transition is suppressed completely, and $\rho$ shows a
linear decrease with temperature down to 25~K. At $T\simeq 22$~K,
$\rho$(T) drops abruptly and reaches zero value by 20~K due to
the onset of superconductivity. The low-$T$ part of $\rho(T)$
measured at various applied fields ($H$) is plotted in the inset
of Fig.~\ref{resistivity} to highlight the variation of
superconducting transition temperature ($T_{C}$) with H. As the
field $H$ increases, the $T_{C}$ decreases gradually.

\begin{figure}
\includegraphics[scale=0.8]{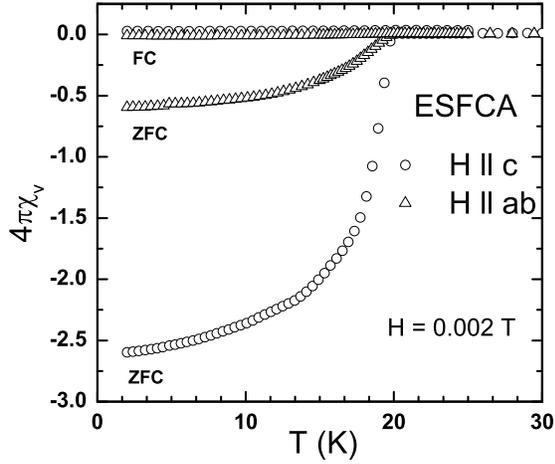}
\caption{\label{ZFC-FC} Temperature dependence of ZFC and FC
magnetic susceptibility of single crystalline ESFCA measured at
an applied field of 0.002 T along two orientations.}
\end{figure}
Beside the transport measurements magnetic measurements are
performed to probe the superconducting transition.
Figure~\ref{ZFC-FC} shows the ZFC and FC dc susceptibility for
the two directions. The field applied is 0.002~T and $\chi$(T) is
plotted in units of $4\pi \chi_{v}$ where $\chi_{v} = -1/4\pi$
indicate complete diamagnetic behaviour. The superconducting
transition at about 20 K is confirmed by the ZFC signal. Here
$\chi_{v}$ for $T\rightarrow 0$ is strongly enhanced for the
H$\parallel$$c$ direction which could be assigned to the
demagnetization effect. Furthermore the finite size of the crystal
in relation to the superconducting penetration depth also
influence the magnetization. The a and b dimension are larger than
the c dimension which strongly increases the demagnetization
factor resulting to an enhanced magnetization for that direction.
Down to 1.8~K, no signature of Eu$^{2+}$ ordering was observed in
the ZFC and FC $\chi$(T) measurements. Nevertheless an AFM
transition superimposed by a large response from a superconducting
transition is not easy to resolve in the SQUID.

\begin{figure}
\includegraphics[scale=1.0]{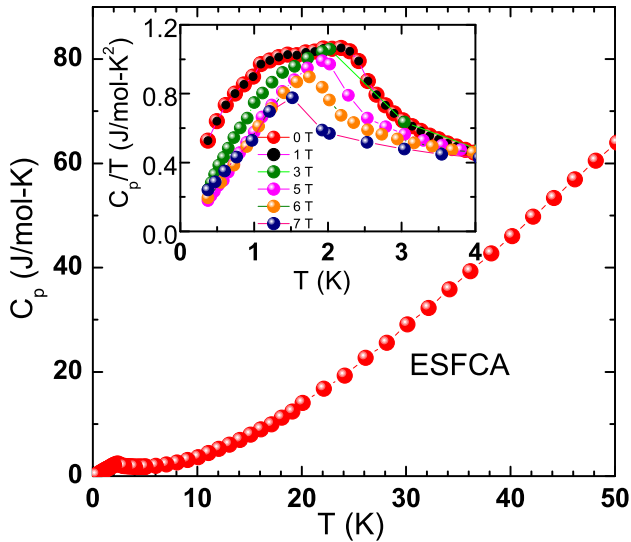}
\caption{\label{specificheat} Temperature dependence
of the heat capacity $C_{\rm p}$ of ESFCA. Inset shows
$C_{\rm p}/T$ vs. $T$ measured at different fields in the low temperature
regime.}
\end{figure}

As a final proof we carried out specific heat measurements on
single crystals, to check whether there exists the heat capacity
jump associate with $T_{C}$ and how far the Eu$^{2+}$ ordering is
suppressed, over the temperature range $0.37~\rm \leq T \leq
50~\rm K$. Figure~\ref{specificheat} shows the $C_{\rm p}$(T)
data measured at zero field. The SC transition is not visible
because the small  jump of $\Delta C_{\mathrm{SC}} \simeq$ 0.8
J/mol-K at $T_{c}\simeq$ 20 K could not be easy to extract from
the phonon dominate specific heat\cite{Zaanen2009}. Nonetheless
the magnetic ordering of Eu$^{2+}$ could be clearly identified. A
sharp anomaly however was observed at about 2~K which can be
attributed to the Eu$^{2+}$ ordering ($T_{\rm N}$). To understand
the nature of the magnetic ordering we measured $C_{\rm p}$(T) at
different applied fields upto 7~T in the low-$T$ regime and
$C_{\rm p}/T$ vs. $T$ is presented in the inset of
Fig.~\ref{specificheat}. With increase in field the height of the
anomaly decreases and $T_{\rm N}$ was found to move towards low
temperatures as is expected for an AFM ordering. Furthermore
integrating $C_{p}/T$ over temperature range 0.37-6 K associated
with the anomaly reveals an entropy gain $\Delta$S=4, which is
23$\%$ of \textit{R}ln8.

\begin{figure}
\includegraphics[scale=1.05]{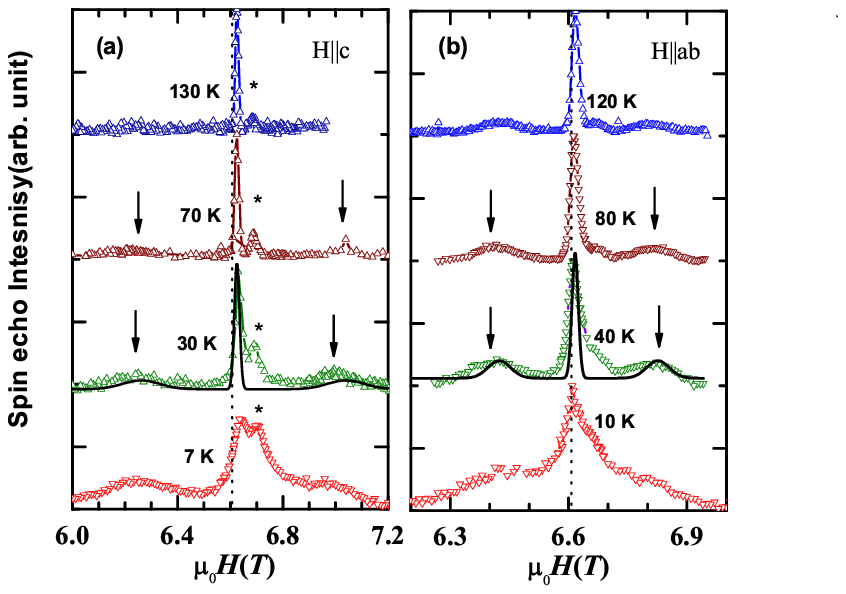}
\caption{\label{spk} $^{75}$As field sweep spectra of ESFCA
measured at various temperatures with external field applied
parallel to the (a) $c$-direction (H$\parallel$$c$) and (b)
$ab$-plane (H$\parallel$$ab$). The down arrows exhibit the
satellite transitions ($\pm 1/2\leftrightarrow\pm3/2$) and the
solid lines are the theoretical fits with the parameters given in
the text. The star mark points to the extra peak close to the
central line for H$\parallel$$c$. Dotted vertical lines indicate
the position of the Larmor field.}
\end{figure}

\subsection{\label{sec:level2}$^{75}$As NMR}
$^{75}$As-NMR measurements are performed on the two different
crystallographic directions (H$\parallel$$c$ and
H$\parallel$$ab$). $^{75}$As-field sweep NMR spectra are shown in
Fig.~\ref{spk}. For both configurations, along with the most
intense central line the spectrum contains extra shoulder-like
features on either side. $^{75}$As has an electric quadrupolar
moment that interacts with the local electric field gradient
(EFG) in the crystal giving rise to the splitting of the NMR
line. Therefore in principle, in case of lower crystal symmetry
system (for example tetragonal and orthorhombic symmetry), one
should see in the $^{75}$As spectra three allowed transitions:
$I_{z}= -1/2 \longleftrightarrow +1/2$ central transition, and
the two $I_{z}= \pm1/2 \longleftrightarrow \pm3/2$ satellite
transitions. Thus the extra shoulders in the experimental spectra
correspond to the first order splitting satellite transitions as
indicated by the downward arrows in Fig.~\ref{spk}. In an attempt
to fit the experimental spectra taking into account both the EFG
and the Knight shift effects in different axes, we find that the
spectra 30~K (H$\parallel$$c$) and 40~K (H$\parallel$$ab$) can be
fitted reasonably well with the parameters (Knight-shift $K_{\rm
c} \simeq -0.55\%$, quadrupolar frequency $\nu_{Q} \simeq 2.82$
MHz, width of central peak $\simeq 145 $ kHz, width of satellite
$\simeq 1050$ kHz, and EFG asymmetry parameter $\eta \simeq 0$)
and ($K_{\rm a,b} \simeq - 0.321\%$, $\nu_{Q} \simeq 1.47$~MHz,
width of central peak $\simeq 228.55$ kHz, width of satellite
$\simeq 600$ kHz, and $\eta \simeq 0$), respectively (see
Fig.~\ref{spk}) These $\nu_{Q}$ values are slightly higher than
that reported for single crystalline SrFe$_{2}$As$_{2}$ compound
\cite{Kitagawa2009}. The central line position is found to be
almost temperature independent or shifting very weakly. At low
temperatures, the NMR line is found to be broaden abruptly. As
seen from the $C_{\rm p}$(T) measurements Eu$^{2+}$ orders
antiferromagnetically below 2~K. Thus our NMR line broadening
possibly arising due to the persistent magnetic correlation while
approaching the Eu$^{2+}$ ordering.
\\
\indent Rather broad satellite transitions are observed in
comparison to the sharp central transition indicating a
distribution of EFG. Due to the doping of Sr and Co, the
disorderness has been inevitably introduced in this system,
resulting a rather distribution of EFG.  In Fig. \ref{spk}(a)
apart from the central and satellite transition another small
peak is observed at around 6.65 T (assigned as $\ast$). The small
peak is originated because of the substitution of 14$\%$ Co at
the Fe site which essentially modify the As neighbors
\cite{Ning2010}. By lowering the temperature the central
transition and the line marked by the $\ast$, broadens and
shifted concurrently to some equal extent. This perhaps indicates
that As in the midst of different nearest neighbors have sensed
the same magnetism. The perceptible line broadening could be
described by two possibilities. First with lowering of the
temperature the inevitable disorderness of this alloy have
sponsored some additional line broadening and the second one
could be associated with Eu$^{2+}$ AFM ordering at low
temperature. Nevertheless, the second argument is more likely as
this supports field dependent specific heat scenario.  Usually
when a system is approaching towards the (AFM/FM) long range
ordering due to the development of internal field associated with
the magnetism the spectra broadens. Being a local probe NMR can
sense this effect well above the ordering.
\begin{figure}
\includegraphics[]{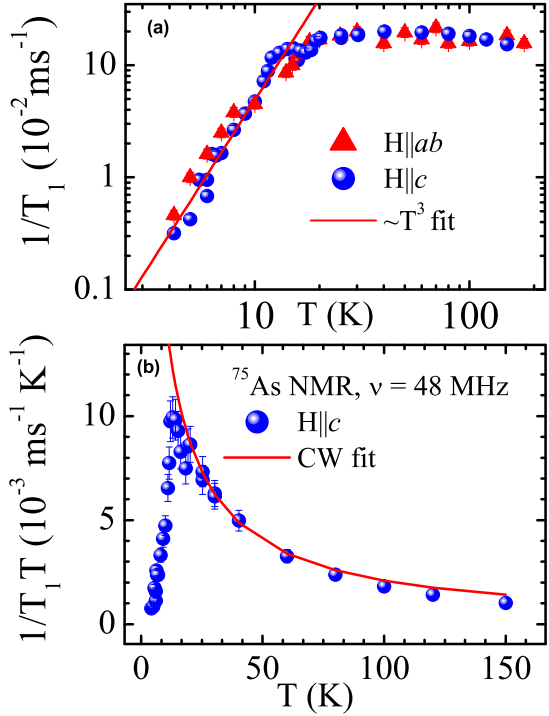}
\caption{\label{spin-lattice} (a) Temperature dependence of the
$1/T_{1}$ for two external field directions H$\parallel$ab and
H$\parallel$c. The solid line represents the $T^{3}$ behaviour.
(b) $1/T_{1}T$ as a function of temperature. The solid line
exhibits Curie-Weiss fit $\frac{C}{T+\theta}$.}
\end{figure}

 Spin-lattice relaxation rate $1/T_{1}$ was measured by
saturating the central line in the two field directions i.e.
H$\parallel$$c$ and H$\parallel$$ab$ by the standard saturation
recovery method. The recovery curves could be fitted consistently
with single $T_{1}$ component using the following equation down
to 18 K,
\begin{equation}
1-\frac{M(t)}{M(\infty)}=0.1e^{-t/T_{1}}+0.9e^{-6t/T_{1}},
\label{exp}
\end{equation}
where $M(t)$ is the nuclear magnetization at a time $t$ after the
saturation pulse and $M(\infty)$ is the equilibrium magnetization.
For $T<18$~K, another short $T_{1}$ component is required to fit
the recovery curves. As seen from Fig.~\ref{spk}, the extra peak
close to the central line become more pronounce at low
temperatures. Since $1/T_{1}$ was measured at the central line
position, we possibly saturate a fraction of that extra peak
which is having a short $T_{1}$ component. Another possibility
could be the low temperature Eu$^{2+}$ AFM ordering. The
temperature dependence of $1/T_{1}$ is presented in
Fig.~\ref{spin-lattice} measured for $H
\parallel ab$ and $H
\parallel c$ directions. For both the directions, the over all
behaviour and magnitude of $1/T_{1}$(T) almost remains same. This
indicates there is no significant anisotropy in the spin dynamics.
At high-$T$s, $1/T_{1}$ is almost $T$-independent. With decrease
in $T$ ($T \lesssim$18~K), $1/T_{1}$ shows a gradual decrease
without showing Hebel-Slichter coherence peak, benchmark for the
$s$-wave type superconductivity. Here it is important to mention
that the absence of Hebel-Slichter peak could be because of the
extended s-wave scenarios, to which s$_{\pm}$ state also belongs
\cite{Parker2008}. The change of slope in $1/T_{1}$ at $T\simeq
18$~K is likely due to the onset of superconductivity and is
consistent with the $\rho(T)$ data at H=7~T which exhibit a sharp
drop nearly at the same temperature (see Fig.~\ref{resistivity}).
In the superconductive regime ($T \leq$ 15~K), $1/T_{1}$ follows
a $T^{3}$ behavior. This possibly is an indication of non
$s$-wave type superconductivity and/or line nodes in
superconductivity gap.

\indent The nuclear spin-lattice relaxation rate,
$\frac{1}{T_{1}}$, is related to the dynamic susceptibility
$\chi_{M}(\vec{q}, \omega_{0})$ per mole of electronic
spins\cite{Johnston2010,Moriya1963} as
\begin{equation}
\frac{1}{T_{1}T} = \frac{2\gamma_{N}^{2}k_{B}}{N_{\rm A}^{2}}
\sum\limits_{\vec{q}}\mid A(\vec{q})\mid
^{2}\frac{\chi^{''}_{M}(\vec{q},\omega_{0})}{\omega_{0}},
\label{t1form}
\end{equation}
where the sum is over wave vectors $\vec{q}$ within the first
Brillouin zone, $A(\vec{q})$ is the form factor of the hyperfine
interactions as a function of $\vec{q}$ in units of Oe/$\mu_{\rm
B}$, and $\chi^{''}_{M}(\vec{q},\omega _{0})$ is the imaginary
part of the dynamic susceptibility at the nuclear Larmor
frequency $\omega _{0}$ in units of $\mu_{\rm B}$/Oe. The uniform
static molar susceptibility $\chi=\chi_{M}^{'}(0,0)$ corresponds
to the real component $\chi_{M}^{'}(\vec{q},\omega _{0})$ with
$q=0$ and $\omega_{0}=0$. According to the mean-field theory,
when the relaxation process is dominated by the
two-dimensional(2D) and three dimensional(3D) AFM fluctuations,
Eq.~\ref{t1form} reduces to $\frac{1}{T_{1}T} \sim
\frac{C}{T+\theta}$ (Curie-Weiss law) and $\frac{1}{T_{1}T} \sim
\frac{C}{(T+\theta)^{0.5}}$, respectively.\cite{Johnston2010} In
the Fig.{\ref{spin-lattice}(b), we have plotted $1/T_{1}T$ vs.
$T$. It increases with decrease in temperature down to 18~K
following a Curie-Weiss (CW) law $1/T_{1}T = \frac{C}{T+\theta}$
in the temperature range 18~K $\leq$ 150~K. The solid line shows
the CW fit with parameters  $C \simeq$ 0.22 (ms)$^{-1}$, and Weiss
temperature $\theta \simeq$ 5~K. We interpret this feature as
arising from two dimensional AFM fluctuations.

The positive Weiss temperature indicates that SC is observed with
the complete suppression of magnetic order. This scenario is
similar to the case of LaFeAs(O$_{1-x}$F$_{x}$), F doped system
where Nakai et. al. observed the positive Weiss temperature
\cite{Nakai2008} and has been predicted that the SC emerges when
a magnetic ordering is suppressed. This tendency is similar to
that in $1/T_{1}T$ of Cu in underdoped
La$_{2-x}$Sr$_{x}$CuO$_{4}$, \cite{Ohsugi-60-1991} where
superconductivity also observed with a positive Weiss temperature.

\section{Conclusions}
The presented results point to a bulk SC with a transition at
$T_{\mathrm{C}}$=20 K at zero field. With field the
$T_{\mathrm{C}}$ is shifted towards lower temperature. The field
dependent $C_{\rm p}/T$ study suggests the AFM ordering of
Eu$^{2+}$ moment at $T_{\mathrm{N}}$=2 K. This AFM ordering is
shifted towards lower temperature from 19 K due to the Sr$^{2+}$
substitution at the Eu$^{2+}$ site in EuFe$_{2}$As$_{2}$.

In order to get a deeper microscopic insight we have performed
$^{75}$As NMR investigations on single crystals of ESFCA in the
external field $H\|c$ and $H\|ab$ directions. $^{75}$As field
sweep NMR spectra broadens with lowering the temperature and the
obtained value of $\nu_{Q}$ is somewhat higher than that has been
obtained for SrFe$_{2}$As$_{2}$ single crystals. This line
broadening phenomena is associated with the low temperature
Eu$^{2+}$ AFM ordering and could be too some extent due to the
disorder.  The spin-lattice relaxation rate $1/T_{1}$ results in
the two crystallographic directions indicate the absence of
substantial anisotropy in the spin dynamics. The temperature
dependence of $1/T_{1}$ exhibits decrement below the SC
transition following a $T^{3}$ behaviour without any
Hebel-Slichter coherence peak. This feature indicate the non s
wave type SC and possibly could be explained within the framework
of s$_{\pm}$ model with impurity effect\cite{Ishida-78-2009}.
Moreover, the strong increase of $1/T_{1}T$ following CW law
arising from two dimensional AFM fluctuations.

\indent NMR line broadening gives the supportive evidence of
Eu$^{2+}$ AFM ordering. On the other hand the drop of $1/T_{1}$
following a $T^{3}$ behaviour below 18 K is a signature of bulk
SC. To illuminate this picture one could have possibly consider
the coupling between the Eu$^{2+}$ moment and the SC. Essentially
Eu$^{2+}$ moment polarizes the conduction electrons (CE), and
they are coupled with the $^{75}$As nuclei via the Fermi contact
interaction, as a result, the line broadening showed up in the
$^{75}$As NMR spectra. On the other hand $T_{1}$ is the measure
of low energy electronic spin fluctuations and in the SC arena
this is modified by the electron pairing. Therefore there is a
possibility of coupling between the SC and the Eu$^{2+}$
magnetism. Furthermore it is important to mention that at $T>$20
K nearly $T$-independent behaviour of $1/T_{1}$ is observed in
both directions. This constant $1/T_{1}$ value possibly indicates,
$1/T_{1}$ is dominated by the relaxation process to the spin
fluctuations of localized Eu$^{2+}$ moments \cite{Yamamoto2008}.



\section*{Acknowledgments} RN would like to acknowledge MPG and
DST India for financial support through MPG-DST fellowship. Work
at G\"{o}ttingen University is supported by DFG-SPP 1458.

\end{document}